# 10

# Sports Business Administration and New Age Technology: Role of AI

**Sahibpreet Singh**[*]

**Dr. Pawan Kumar**[**]

## INTRODUCTION

Sports is a booming industry apart from just being a form of exercise. Exchange of billions of dollars from various countries, players, trainers, officials is involved. Sports business administration is the control of sport related undertakings and establishments such as teams, leagues, tournaments, facilities and merchandising. Also referred to as sports management; sports business management encompasses areas like sales promotion, monetary accounting for businesses based on sports concepts job specializations in this field include the fields that involve: Marketing managers who focus on promoting uses for sport-related goods or services; Executive Directors who handle all of the operations on behalf of any specific company in their jurisdiction since there's no other definition similar to it.[1]

Sports law encompasses various laws applied in relation to different aspects of sporting activities such as contract enforcement rules; torts, intellectual property rights, competition law, employment legislation, governance provisions, dispute resolution mechanisms within a particular jurisdiction across several jurisdictions for the betterment of the society.[2] Sports law does not exist as a single unitary body but comprises numerous laws and regulations which can vary widely depending upon sport-specific authorities geographic locations which may be country-wide or provincial/statewide (i.e., federal system) used by those governing bodies concerned with each sport or country involved. Sports law can be divided into two main categories:

    i.     lex sportiva and
    ii.    lex ludica.

---

\*      LLM (2023-24), Department of Laws, Guru Nanak Dev University, Amritsar.

\*\*    Professor, Department of Laws, Guru Nanak Dev University, Amritsar

1.     MASSIMO VALERI, CORPORATE SOCIAL RESPONSIBILITY AND REPORTING IN SPORTS ORGANIZATIONS (2019), http://link.springer.com/10.1007/978-3-319-97649-5 (last visited Feb 11, 2024).

Lex sportiva refers to the rules and regulations that are specific to sports. They are created and enforced by sport's governing bodies. For example, IOC[3], FIFA[4] and BCCI[5]. Lex ludica refers to the general laws and principles that apply to sports. These are derived from national and international legal systems. Primarily from constitutional law, criminal law, civil law and human right laws.[6]

**LEGAL ASPECT OF SPORTS FRANCHISING, LICENSING AND MERCHANDISING**

**Sports Franchising**

Sport franchising is a kind of business model where an independent body referred to as a franchisee is permitted by a league or association to run a team or club under its name and identity. That also includes payment of sums for obtaining the right to use the brand name as well as agreement on certain rules established by the franchise holder. The franchisor supports a franchisee with training in marketing as well as other services. In India there are IPL[7], ISL[8], and PKL[9] teams among others.

The benefits of sports franchising include:

  i.   Allows for expansion into new markets gaining access to more consumers without investing much capital.

---

2.     Sports Law: its History and Growth and the Development of Key Sources | Legal Information Management | Cambridge Core, https://www.cambridge.org/core/journals/legal-information-management/article/abs/sports-law-its-history-and-growth-and-the-development-of-key-sources/277C0C4BF18761696AE39F21791762A8 (last visited Feb 11, 2024).
3.     International Olympic Committee.
4.     Federation Internationale de Football Association.
5.     Board of Control for Cricket in India.
6.     Ken Foster, *Lex Sportiva and Lex Ludica: The Court of Arbitration for Sport's Jurisprudence, in* Lex Sportiva: What is Sports Law? 123 (Robert C.R. Siekmann & Janwillem Soek eds., 2012), https://doi.org/10.1007/978-90-6704-829-3_7 (last visited Feb 11, 2024).
7.     Indian Premier League.
8.     Indian Super League.
9.     Pro Kabaddi League.

ii.    Franchisee can benefit from using the goodwill of the franchisor who has already created a remarkable reputation in addition to bringing in expert knowledge and connections.

iii.   It is a competitive sports product which generates revenue for instance from broadcasting rights sponsorships ticket sales and merchandising.[10]

The challenges of sports franchising include:

a)     It requires the franchisor to maintain quality control and consistency across all the franchises, and ensure that they comply with the rules and regulations of the league or organization.

b)     It requires the franchisee to adhere to the terms and conditions of the franchise agreement, and share a portion of its profits with the franchisor.

c)     It may create conflicts and disputes between the franchisor and the franchisee, or among the franchisees, over issues such as revenue sharing, territorial rights, performance standards, and termination clauses.[11]

The legal aspects of sports franchising involve the following:

1) The franchise agreement, which is the contract that governs the relationship between the franchisor and the franchisee, and specifies their rights and obligations. It shall include intellectual property rights of franchisor, the standards for quality, termination clauses and other details regarding how long or wide it should be. Additionally, this agreement should account for the Indian Contract Act 1872 along with Foreign Exchange Management Act of India 1999. Acts relating to competition like Competition Act 2002 are law that must be complied with by any enterprise engaging in franchising including Goods and Services Tax Act 2017 or applicable legislation related to taxation.

2) The intellectual property rights, which are the rights that the franchisor and the franchisee have over the trademarks, logos, names, slogans, and other distinctive elements of the franchise. In most cases, franchisor owns intellectual property rights instead of franchisee but they grant franchises through a license arrangement during particular periods and for certain causes. Nevertheless, such proprietors ought to register these trademarked properties under Trademarks Act 1999 so that later they could enforce them civilly or criminally against infringement dilution third parties whose misuse can happen.

3) The regulatory compliance, which is the obligation of the franchisor and the franchisee to follow the rules and regulations of the sports league or organization, as well as the relevant authorities and bodies, such as the Ministry of Youth Affairs and Sports, the Sports Authority of India, the National Anti-

---

10.    The Pros and Cons of Investing in a Sports Franchise, INVESTOPEDIA, https://www.investopedia.com/articles/fundamental-analysis/12/pros-cons-investing-sports.asp (last visited Feb 11, 2024).

11.    Jeff Harris, Luc Ryu & Marco D'Elia, *VALUATION OF SPORTS TEAMS AND FRANCHISES.*

Doping Agency, and the National Sports Federations. Both parties involved in franchising have an obligation to adhere to the ethical values and standards of conduct that apply in sports sector including match fixing doping corruption discrimination etc.

**Sports Licensing**

Licensing in sports is an agreement where a sports enterprise like a league, team, club or player grants permission to use their name or logo among other intellectual property to another entity called licensee for specific reasons such as making selling goods, creating media content and organizing events. The licensee pays some fee and agrees to certain terms set by the licensor. Licensor keeps ownership of intellectual property and controls its quality through monitoring the performance of licensee.[12] For example licensing of IPL team logos & names with several merchandise manufacturers & retailers in India.

The benefits of sports licensing include:

i. In this arrangement, the licensor can make some more money and sell his/her ideas to a wider audience without putting in too much money into it.

ii. The licensee benefits from the following and goodwill of the licensor while offering a product or service that is different from others.

iii. A loyal and active fan base for sports team is built which increases its brand value and awareness.[13]

The challenges of sports licensing include:

a) It requires the licensor to carefully select and monitor the licensee, and ensure that it maintains the quality and reputation of the intellectual property.

b) It requires the licensee to comply with the terms and conditions of the license agreement, and pay the fees and royalties to the licensor.

c) It may create conflicts and disputes between the licensor and the licensee, or among the licensees, over issues such as the scope and duration of the license, the fees and royalties, the intellectual property rights, the quality

12. Yellowbrick, *The Definitive Guide to Sports Licensing*, YELLOWBRICK (2023), https://www.yellowbrick.co/blog/sports/the-definitive-guide-to-sports-licensing (last visited Feb 18, 2024).
13. Main Goals & Objectives in Sports Licensing, https://smallbusiness.chron.com/main-goals-objectives-sports-licensing-37071.html (last visited Feb 18, 2024).

standards, the dispute resolution mechanisms, and the termination clauses.[14]

Legal aspects of sports licensing involve the following:

1) The license agreement, which is the contract that governs the relationship between the licensor and the licensee, and specifies their rights and obligations.

2) The intellectual property rights, which are the rights that the licensor and the licensee have over the trademarks, logos, names, images, and other distinctive elements of the sports entity. It is the licensor who usually owns IP rights for a specified period or purpose.[15] To safeguard their IP against infringement, dilution or misappropriation by third parties, the licensor and licensee must register it. Later on enforce it through civil or criminal remedies.

3) The regulatory compliance, which is the obligation of the licensor and the licensee to follow the rules and regulations of the sports league or organization, as well as the relevant authorities and bodies, such as the Ministry of Youth Affairs and Sports, the Sports Authority of India, the National Anti-Doping Agency, and the National Sports Federations.

**Sports Merchandising**

It is a business activity where a sports entity manufactures and sells items or services connected to its name, logo, image or other intellectual property. Sports entity includes a league, team, club, or an individual player. They can produce the merchandise and sell. The same can be licensed to third-party manufacturers.[16] Additionally they can cooperate with other brands and products by means of co-branding. Examples include IPL team jerseys and caps and mugs, etc.

Certain benefits include the following:

i.      It allows the sports entity to generate additional income and exposure from its intellectual property, and diversify its revenue streams.

ii.     More so sport entity can develop its brand value & recognition while creating distinct identity & image in the market place.

---

14. Navigating Legal Challenges in Sports Licensing and IP - Yellowbrick, https://www.yellowbrick.co/blog/sports/navigating-legal-challenges-in-sports-licensing-and-ip (last visited Feb 18, 2024).

15. Sports and Intellectual Property, https://www.wipo.int/en/web/sports/ (last visited Feb 18, 2024).

16. Global Licensed Sports Merchandise Market Size Report, 2030, https://www.grandviewresearch.com/industry-analysis/licensed-sports-merchandise-market-report (last visited Feb 18, 2024).

iii.    Moreover sport entity can connect & interact with fans & consumers as well as increase their loyalty & engagement levels.[17]

The challenges include:

a)    It requires the sports entity to invest in the production, distribution, and marketing of the merchandise, and ensure its quality and availability.

b)    It requires the sports entity to protect its intellectual property rights from infringement, dilution, or misuse by third parties, and avoid any conflicts or disputes with its licensees or co-branding partners.

c)    It requires the sports entity to comply with the applicable laws and regulations, such as the Consumer Protection Act, 2019, the Legal Metrology Act, 2009, the Bureau of Indian Standards Act, 2016, and the Goods and Services Tax Act, 2017.

The following important legal aspects:

1) The intellectual property rights, which are the rights that the sports entity has over the trademarks, logos, names, images, and other distinctive elements of its merchandise. The sports entity should register its intellectual property rights under the relevant laws. To keep track of the market and prevent use of its intellectual property without the consent from third parties or counterfeits thereof the sports entity can do this.

2) The contractual agreements, which are the contracts that the sports entity enters into with its licensees, co-branding partners, manufacturers, retailers, or consumers, and specify their rights and obligations.

3) The consumer protection, which is the obligation of the sports entity to ensure that its merchandise is safe, genuine, and of good quality, and that it does not deceive, mislead, or harm the consumers.

**PROTECTION OF INTELLECTUAL PROPERTY RIGHTS IN SPORTS**

IPR is quite essential for the development and growth of this sector today. These enable the creation, protection, and exploitation of various sports-related assets. Various designs, logos, trademarks, names, images, etc are proctected. Nonetheless, these IPRs are also subject to many challenges and threats within the sport field including infringements, counterfeiting piracy ambush marketing unfair competition etc.[18]

**Importance of protecting and enforcing IPRs in sports**

i.    Safeguarding Intellectual Property protects the ownership over unique valuable assets thereby preventing unauthorized use by third parties.

---

17.    Author    Megha    Singh,    *Benefits    of    Sports    Marketing*,    (Dec.    29,    2023), https://www.themediaant.com/blog/benefits-of-sports-marketing/ (last visited Feb 18, 2024).

18.    Aswathy Sujith, *SPORTS AND INTELLECTUAL PROPERTY RIGHTS – AN OVERVIEW ON THE INDIAN STANDARDS*, 2.

ii.   Sports entities could gain income from the assets by licensing, selling or transferring them to others.

iii.  Sports entities have the capacity of adding value to their brand and gaining recognition while creating a lasting and engaged fan base through delivering differentiated or attractive sports product/service and communicating its image to public.[19]

iv.   Investment in R&D plus adoption of new technologies enables sports organizations to promote innovation and creativity as well as enhance quality and performance of sport products/services.[20]

v.    It enables the sports entities to contribute to the social and economic development of the society, by creating jobs, developing infrastructure, promoting sports culture, and supporting sports education and training.[21]

**Challenges of protecting and enforcing IPRs in Sports**

a)   The sports sector is dynamic and complex, involving multiple stakeholders, activities, and markets, across different jurisdictions and cultures, which creates legal uncertainty and diversity in the interpretation and application of IPRs.

b)   The sports sector is highly competitive and lucrative, attracting many opportunistic and malicious actors, who seek to exploit the popularity and goodwill of the sports entities, and to benefit from their assets, without their consent or compensation.

c)   The sports sector is vulnerable and exposed, due to the widespread and intensive media coverage, the rapid and easy dissemination of information and content, and the high public interest and demand, which facilitate the infringement and violation of IPRs.

d)   The sports sector is evolving and transforming, due to the emergence and adoption of new technologies, platforms, and formats, such as digital media, social media, streaming services, and e-sports, which pose new challenges and threats to the protection and enforcement of IPRs.

## LEGAL FRAMEWORK AND BEST PRACTICES FOR SPORTS GOVERNANCE AND COMPLIANCE

Sports governance is closely related to compliance because it touches on the structure, management, oversight over athletics organizations. Governance in sport mainly encompasses establishment as well as application of regulations, policies together with procedures that direct decision-making process within sporting agencies among other stakeholders. Sports Compliance is the act of sticking to and conforming with the existing regulations legislations, standards, and ethical values that govern the activities of both sports organisations and its members. The chief goals of sports governance and compliance are to ensure transparency, accountability, integrity, and efficacy within sports establishments as well as to guard athletes, fans, sponsors and so on. A number of factors influence sports governance and compliance such as legal autonomy of sport organizations; commercialization of sport; internationalization of sport; diversity in stakeholders; new challenges and opportunities for the sporting industry. This results in complex dynamic environment where sport organizations must implement good governance and comply practices that are relevant to their specific context and objectives.

---

19.   Kyle D. S. Maclean, *Sales and Revenue Generation in Sport Business by David J. Shonk and James F. Weiner. 2021*, 22 J Revenue Pricing Manag 231 (2023).

20.   Jonas Hammerschmidt et al., *Sport Entrepreneurship: The Role of Innovation and Creativity in Sport Management*, Rev Manag Sci (2023), https://doi.org/10.1007/s11846-023-00711-3 (last visited Feb 18, 2024).

21.   Sean Ennis, *Sport and Its Role and Contribution to Society and Economic Development*, in Sports Marketing: A Global Approach to Theory and Practice 9 (Sean Ennis ed., 2020), https://doi.org/10.1007/978-3-030-53740-1_2 (last visited Feb 18, 2024).

**Sources and Principles of Sports Law**

**Sports law** is a multidisciplinary and international field of law that covers various legal areas and sources that are applicable to sporting activities and organizations. Sports law can be divided into two categories namely public sports law and private sports law.

**Public sports law** refers to the regulations set by state organs like national, regional, and local governments, courts, or administrative institutions. Public sports laws govern public aspects such as recognition and funding of sport organizations, safety at sporting events, anti-doping/anti-corruption measures at these events, tax incentives for sport activities etc; human rights and fundamental freedoms of participants in sport.

Under **private sports law**, the sporting organizations themselves which include international, national and regional sports federations, leagues, clubs and associations create and enforce rules and regulations. Private sports law is about how particular aspects of sport are run as in who can become a member of a sport organization or how a board should be constituted; the regulation of athletes' participation; the rules of games; contractual relationships in sports business; IP rights protection in sport and licensing issues concerning them; dispute resolution mechanisms within this field.

A similarity between public and private sports laws is that they are founded on general legal principles like rule of law, separation of powers, due process of law, equality before the law, proportionality principle, legal certainty principle and good faith.[22] Also public as well as private sports laws incorporate international legal norms that include autonomy for sport institutions as well as their self-regulation by example where governing relations between various

---

22. Dimitrios Panagiotopoulos, *GENERAL PRINCIPLES OF LAW IN INTERNATIONAL SPORTS ACTIVITIES AND LEX SPORTIVA* (2015).

FIFA members are concerned; solidarity among those involved with ensuring social responsibility around sport events; equity during competitions – no unfairness towards/ favoring either side (which usually occurs through match fixing); protecting rights & interests of athletes too such statements like "the University Athletes Association needs to be registered"; respecting diversity & singularity inherent in every athletic practice.[23]

**Organisational Structure and Functioning Of Sports Bodies**

Sports regulatory bodies are characteristically concerned with their organisational structure and functioning. The way sports organisations are designed and composed, that is division and distribution of roles, responsibilities and powers among the different organs, units or members of the organisation is referred to as organisational structure. On the other hand, organisational function refers to how sports organizations run or operate such as when they adopt policies and strategies in order to achieve their objectives.[24]

The type, level, and scope of the organization determine the organizational structure and functioning of sports bodies. For instance, an international federation like International Olympic Committee (IOC) or International Cricket Council (ICC) would have a different organizational structure and way of operation from a national federation like Indian Olympic Association (IOA) or Board for Control Cricket in India (BCCI), while regional or local federation like Punjab Olympic Association (POA) or Amritsar District Cricket Association (ADCA) would exhibit its own features. Similarly, The Indian Premier League (IPL) or the Indian Super League (ISL), for example, is organised and functions differently from a sports club like Mumbai Indians or Kerala Blasters or a sports association such as Professional Golfers Association of India (PGAI) or All India Tennis Association (AITA).[25]

Nevertheless, each type, range and magnitude of the sports organization has some common elements that denote its organizational structure and functioning. These include:

i.   The constitution or statutes of the sports organisation encapsulate its name, objectives, values and visions as well as its membership, structure, governance and administration.

ii.  General assembly of congress of the sport organization is the highest body within this organization which is formed by representatives of the members who constitute it, possesses powers to elect governing bodies, approve budget and accounts; amend its statute/constitution; make any other decisions on matters dealing with general interest to that entire body.

---

23.   Foster, *supra* note 3.
24.   Sandalio Gomez, Magdalena Opazo & Carlos Marti, *Structural Characteristics of Sport Organizations: Main Trends in the Academic Discussion*, SSRN JOURNAL (2008), http://www.ssrn.com/abstract=1116226 (last visited Feb 18, 2024).
25.   Diksha Garewal Das, *Sports Organisations and Governance in India*, 10 (2020).

iii. The highest executive and representative organ of the sports organisation, which is the governing body or executive committee, constituted by elected or appointed members of the organization while having powers to implement resolutions of general assembly, oversee the operations as well as represent it externally.

iv. Sporting organizations' heads who are presidents or chairpersons are usually chosen through election by a general assembly or appointment by a governing body with authority to lead over the meetings of general assemblies and governing bodies, control organizations functioning and assign tasks to other members of organs in those organizations.

v. The secretary-general or chief executive officer of the sports organisation, who is the chief administrative and operational officer of the organisation, appointed by the governing body or the president, and who has the power to oversee the daily management and coordination of the organisation, implement the policies and strategies of the organisation, and report to the governing body and the president.

vi. The specialised and advisory organs of sports organisations that consist of committees or commissions created either by the president himself, president's governing body or congress which he has powers to provide support in various spheres such for example finance, development audit, legal, medical ethics technical athletes women youth etc.

vii. The sport organisation's professional and qualified personnel, who are the staff or workers of the organisation, recruited by the president or the secretary-general and possessing authority to implement and provide all programs and services of an organization such as commerce, educational, administrative, communication sponsoring functions etc.

Various legal and regulatory requirements and standards govern how sports bodies operate. Sports entities, both private and public law, have to adhere to the set regulations.[26] For example, in India non-governmental sports associations must comply with provisions of the Indian Societies Registration Act, 1860; the Indian Trusts Act, 1882; and the Indian Income Tax Act, 1961, among others. There is also a National Code for Good Governance in Sports, 2017 (Draft), which mandates minimal good governance norms that recognised Government of India-funded sports organizations must adopt or face loss of such recognition including but not limited to: eligibility and disqualification criteria for office-bearers; term limits and cooling-off periods for board members; representation and participation of athletes as well as women in decision-making process; transparency/ accountability in financial management; internal grievance redressal/ dispute resolution mechanisms.[27]

---

26. Jitendra Choudhary & Jayat Ghosh, *GOVERNANCE OF SPORTS IN INDIA* (2013).

Apart from the legal and regulatory requirements and standards, sports organizations also need to follow good practices and suggestions for organisational structure and operation of sports bodies at national and international levels. For example, Indian sports organizations can take into account recommendations of various committees or commissions constituted by the Government of India or Supreme Court of India to study issues related to sport governance and management like Mukul Mudgal Committee[28], Justice R.M. Lodha Committee[29], Olympic Task Force[30]. The same way Indian sporting organizations can also look up for best practices and guidelines from worldwide sporting bodies like IOC, ICC, WADA etc that have come up with many codes, norms and frameworks relating to organizational structures as well as operations of such entities including Olympic Charter[31], ICC Constitution[32], WADA Code[33], Basic Universal Principles of Good Governance in Olympic & Sports Movement[34].

The subsequent are the best practices and guidelines to follow when structuring and functioning of sports bodies:

1) To ensure the democratic and participatory governance of sports organisations, by involving and consulting the various stakeholders of the sports sector, such as the athletes, the coaches, the officials, the fans, the sponsors, the media, the civil society, and the government, in the decision-making and policy-making processes of the organisation.

---

27. Draft National Code for Good Governance in Sports.pdf, https://yas.nic.in/sites/default/files/Draft%20National%20Code%20for%20Good%20Governance%20in%20Sports.pdf (last visited Feb 18, 2024).

28. Team Sportstar, *Mukul Mudgal Elected Head of FIFA's Governance, Audit and Compliance Committee*, SPORTSTAR (2021), https://sportstar.thehindu.com/football/mukul-mudgal-elected-head-of-fifas-governance-audit-and-compliance-committee/article34627495.ece (last visited Feb 18, 2024).

29. BCCI – Lodha Committee – IAS4Sure, https://www.ias4sure.com/wikiias/gs2/bcci-lodha-committee/ (last visited Feb 18, 2024).

30. Olympic Task Force recommends steps to improve sports in India | Sport-others News - The Indian Express, https://indianexpress.com/article/sports/sport-others/olympic-task-force-recommends-steps-to-improve-sports-in-india-4793854/ (last visited Feb 18, 2024).

31. Olympic Charter, https://olympics.com/ioc/olympic-charter (last visited Feb 18, 2024).

32. ICC Constitution, ICC - INTERNATIONAL CHAMBER OF COMMERCE, https://iccwbo.org/icc-constitution/ (last visited Feb 18, 2024).

33. The World Anti-Doping Code, https://www.wada-ama.org/en/what-we-do/world-anti-doping-code (last visited Feb 18, 2024).

34. Universal principles for integrity, https://olympics.com/ioc/integrity/universal-principles-for-integrity (last visited Feb 18, 2024).

2) To ensure the independent and impartial governance of sports organisations, by avoiding and preventing any conflicts of interest, undue influence, or interference from any external or internal sources, and by establishing and enforcing clear and transparent rules and procedures for the election, appointment, and removal of the members and organs of the organisation.

3) To ensure the efficient and effective governance of sports organisations, by defining and clarifying the roles, responsibilities, and powers of the different members and organs of the organisation, and by establishing and implementing appropriate mechanisms for the coordination, communication, and collaboration among them.

4) To ensure the accountable and transparent governance of sports organisations, by disclosing and reporting the relevant information and data about the activities, performance, and finances of the organisation, and by establishing and applying adequate mechanisms for the monitoring, evaluation, and audit of the organisation.

5) To ensure the sustainable and responsible governance of sports organisations, by aligning and integrating the goals and objectives of the organisation with the social, environmental, and economic dimensions of the sports sector, and by establishing and promoting the ethical values and principles of the organisation.

## TAXATION POLICIES AND INCENTIVES FOR SPORTS ACTIVITIES

Sports being a recreational activity, also plays out to be a significant contributor to the economy.[35] Various stakeholders get employment through sports. Additionally, sports support health care promotion, social integration cultural diversity and national pride.[36] As such there should be supportive taxation policy system in place towards the sporting activities within nation states.

While Income Tax Act 1961 governs taxation of sports activities in India; Goods and Services Tax (GST) Act 2017 controls this field. The Income Tax Act subjects different sport organizations including teams like associations as well as clubs. The GST Act provides that goods and services relating to sporting events will be subject to GST such as tickets for entry into stadia broadcasting rights fees paid by media houses sponsorship fees related to various merchandise sold at stadia coaching fee earned among others.

---

35. Study on the contribution of sport to economic growth and employment in the EU.
36. Guest Author, *The Economic Impact of the Sports Industry: Driving Growth, Job Creation, and Revenue Generation*, LA PROGRESSIVE (2023),
    https://www.laprogressive.com/sponsored/economic-impact-of-the-sports-industry
    (last visited Feb 18, 2024).

Sports activities and events can be promoted in certain ways on the basis of provisions contained in The Income Tax Act as follows:

i.    Since it is the guiding principle for taxation of sports concerning different countries, section 10(39) of ITA, India makes such acts taxable under certain conditions for any event held in our country which has been certified by an international body and participated by more than two nations as well as declared by Central Government. This exemption is meant to promote international sports in India and attract players and spectators from other countries.

ii.    Section 10(17A) of the Income Tax Act exempts any income, in any form that is awarded to an individual by either the Central Government or State Government or by any other institution established in India to reward outstanding performance within fields of activity including sports.

iii.    Section 80G gives provision for deductions of donations made towards specified funds which are approved by Central Government for promoting sports in India.

iv.    The Income Tax Act provides for a special tax rate of 20% (plus surcharge and cess) under section 115BBA on the income earned by non-resident sportsmen or sportswomen, or non-resident sports associations or institutions who participate in any game or sport in India as well as make any contributions such as articles, advertisements etc., to an Indian publication. This is meant to facilitate taxation of foreign sports persons and entities simply and avoid disputes due to double taxation.

Some of these incentives and exemptions are there in the GST Act concerning sports activities and events:

a)    The GST rate on admission to sports events is 18%, which is lower than the standard rate of 28% applicable to entertainment events such as cinema, theatre, and amusement parks.

b)    The GST rate on coaching services provided by independent coaches or trainers is 18%, which is lower than the rate of 28% applicable to coaching services provided by commercial coaching or training centres.

c)    The GST rate on supply of sports goods is 12%, which is lower than the rate of 18% applicable to most other goods.

d)    The GST Act exempts the supply of services by way of participation in a sports event organised by a national sports federation, or its affiliated federations, where the participating team or individual represents any district, state, zone, or country.

e)    The GST Act also exempts the supply of services by way of sponsorship of sporting events organised by a national sports federation, or its affiliated federations, or a sports body registered under any law for the time being in force.

On the other hand, there are difficulties in regards to tax systems that surround sporting activities:

1) The lack of clarity and uniformity in the definition and classification of sports activities and events, and the distinction between sports and entertainment, which may lead to confusion and disputes among the taxpayers and the tax authorities.[37]

2) The multiplicity and complexity of taxes and compliances, especially for cross-border transactions and payments, which may increase the cost and burden for the sports persons and entities, and discourage foreign investment and participation in sports.

3) The absence of specific tax incentives and exemptions for certain segments and aspects of sports, such as sports infrastructure, sports education, sports research, sports tourism, sports medicine, sports technology, and sports innovation, which may hamper the holistic and integrated development of sports in India.[38]

Thus, it is required that the taxation policy regarding sports activities be reformed. This will make sure that this policy aligns itself with the best practices as well as international standards. Some possible suggestions and recommendations include:

i. To clear out a coherent elucidation along with categorization of sports activities/events across various tax legislations/ guidelines.

ii. To rationalize tax rates on all sporting activities/events while simplifying their compliances while also introducing a single-window clearance system for cross-border transactions/payment disputes.

iii. To introduce specific tax incentives and exemptions tied to performances/outcomes of sports personnel/entities for Sports Infrastructure, Sports Education, Sports Research, Sports Tourism, Sports Medicine, Sports Technology, Sports Innovation etc.

iv. To increase public-private partnership as well as corporate social responsibility in sports sector and encourage stakeholders such as firms, NGO's press and civil society to contribute and participate on issues of sport development.[39]

Through these means the tax policy for sporting activities/events can be made more conducive.

---

37. Taxing International Sporting Events—the India Perspective, https://news.bloombergtax.com/daily-tax-report-international/taxing-international-sporting-events-the-india-perspective (last visited Feb 18, 2024).
38. Sports infrastructure: Transforming the Indian sports ecosystem.
39. Vijay Krishnamurthy, *Public-Private Partnership (PPP) in Sports Sector - A Scan of Extant Literature with a Special Focus on Odisha State* (2021).

**DISPUTE RESOLUTION MECHANISMS AND ARBITRATION IN SPORTS**

Disagreements amongst different stakeholders are inevitable. They can arise because of difference in opinion. Such disputes originate from contract wrangles, disciplinary actions, doping offences or any criticism relating to qualification. Selection criteria, governance issues, intellectual property rights, corruption are among them.[40] Another thing is to have mechanisms that work efficiently and effectively in addressing sports disputes without having to go through a long torturous process which also brings about expensive litigation processes. The success of resolving sport's dispute could affect their lives, image and financial aspects of the parties involved as well as integrity and growth of sport itself.[41] This is where alternative dispute resolution comes up which is one of the most utilized and accepted modes for solving sport's related issues globally is arbitration. One or a group may be made of arbitrators who are experts either in law or sporting field hence this method sometimes called private adjudication since it operates outside normal tribunals. Arbitration decisions are final and enforceable by any court in the world depending on its jurisdictional position.[42]

Arbitration has a number of advantages over litigation in sports disputes. These include:

1.  **Neutral Forum**: Through arbitration, neutral forums, and processes can be provided for solving cross-border or international sports conflicts which may involve local courts and laws that may be biased or unfamiliar.

2.  **Flexible**: In relation to the preferences and needs of the parties involved, arbitration allows them to choose their arbitrators, the governing law, language, rules of evidence and procedure of arbitration.

3.  **Expert Decision Making**: The arbitration facilitates appointments of arbiters who are knowledgeable with regard to specialized legal field or sport ensuring a decision is well-informed as well as high-quality.

4.  **Secrecy**: By this means it is possible to shield both the participants and the proceedings themselves from publicity and media scrutiny by keeping them private while choosing instead an arbitration process than a lawsuit.

---

40.   Suhani Dhariwal, *Arbitration In Sports in India - Court of Arbitration for Sport (CAS)*, WRITINGLAW (Jul. 17, 2023), https://www.writinglaw.com/arbitration-in-sports-in-india/ (last visited Feb 18, 2024).

41.   Sports Dispute Resolution In India - Sport - India, https://www.mondaq.com/india/sport/1177320/sports-dispute-resolution-in-india (last visited Feb 18, 2024).

42.   Arbitration in the Realm of Sports in India: An Analysis - Chanakya Centre for Alternative Dispute Resolution, https://ccadr.cnlu.ac.in/blog/arbitration/arbitration-in-the-realm-of-sports-in-india-an-analysis/ (last visited Feb 18, 2024).

5. **No Appeals Possible**: Lastly, through arbitration one is able to get final decisions that have binding effect on court executions without any opportunity for appealing them apart from very restricted grounds only.[43]

The most significant and influential body in sports arbitration is the CAS or the Court of Arbitration for Sport, established by IOC (International Olympic Committee) in 1983 and which can be found in Lausanne, Switzerland.[44] The court is independent and specializes in sports matters enabling it to resolve them through arbitration or mediation as requested by parties or referred from other sports bodies. The CAS has three divisions:

  i.    the Ordinary Arbitration Division,

  ii.   the Appeals Arbitration Division, and

  iii.  the Ad Hoc Division.

The CAS also has a number of regional offices and ad hoc tribunals across the world. It deals with sports disputes arising out of the statutes or regulations governing the respective sporting federations or associations, or out of specific arbitration agreements between the parties involved.[45] The following sports disputes that can be brought to CAS:

a)   Disputes relating to the application or interpretation of the rules of sports federations or associations, such as eligibility, selection, qualification, ranking, doping, discipline, sanctions, etc.

b)   Disputes arising from contractual relationships between the parties involved in sports, such as employment contracts, sponsorship contracts, transfer contracts, broadcasting contracts, etc.

c)   Disputes arising from tortious liability or other civil liability in sports, such as personal injury, negligence, defamation, etc.

CAS applies a set of regulations known as Code of Sports related Arbitration. It governs all arbitration proceedings conducted at CAS. Moreover, it applies substantive law chosen by the parties themselves or where an alternative law is not determined, then it will apply according to its domicile state federation or association, failing this arbitrators choose what they find suitable. The CJEU also takes into account laws and regulations made by sporting federations/associations including general principles of law/rule of natural justice.[46] CAS arbitration procedures are usually done in a way that is private and fast, with the aim of having an outcome within reasonable time. CAS arbitration awards usually bind the parties to it finally and can be enforced in the courts of any country that is a participant to New York Convention on Recognition and enforcement of Foreign Arbitral Awards. CAS arbitration awards may only be contested before Swiss Federal Tribunal on very few grounds such as absence of jurisdiction, violation of due process or excessive powers. The CAS has been instrumental in enhancing sports law practice and jurisprudence through resolving numerous sports disputes covering various sports, different countries from different regions.[47] Additionally, CAS has contributed towards promotion as well as protection of sport values which include solidarity, integrity, fair play and respect. Most international federations for various games like the IOC, International Paralympic Committee (IPC), World Anti-Doping Agency (WADA) or United Nations have recognized this body.[48] However, there are several challenges and criticisms faced by it, such as:

i.   **Diversity and independence**: The ICAS has been accused of being a monopoly of International Federations supported by the IOC having his headquarters in Switzerland that appoints almost all arbitrators, mainly from Europe and North America, who form the bench for CAS arbitral tribunals. This is likely to raise questions about neutrality and representativeness of CAS as well as the judges themselves.

ii.  **Transparency and accountability**: Generally there are allegations that CAS is too secretive and non-transparent because it does not publish its awards/proceedings on websites, or allow public access to hearings or participation by public. The latter may lead to a credibility gap concerning the CAS's authority as well as decision-making.

iii. **Accessibility and affordability**: Some athletes have challenged the Court of Arbitration for Sport because it is expensive even when individual athletes or small sports organizations cannot afford legal representation before CAS which they might need when filing their claims or responding to them respectively. This would lead to an unfair situation where one party is disadvantaged at the expense of another.

iv.  **Consistency and coherence**: It has also been called into question because some times there are different judgments/contradictory decisions over similar issues arising on different occasions depending on arbitrators, litigants, facts, or jurisprudence. Such may jeopardize the reliability and stability of sports law and jurisprudence.

---

43.   The Case for Alternative Dispute Resolution in Sports | Sports Litigation Alert, https://sportslitigationalert.com/the-case-for-alternative-dispute-resolution-in-sports/ (last visited Feb 18, 2024).

44.   Home, (2024), https://www.tas-cas.org/en/general-information/index/ (last visited Feb 18, 2024).

45.   Clifford J Hendel, *Jurisdiction of the CAS – The Basics* (2017).

46.   CAS_Code_2020__EN_.pdf, https://www.tas-cas.org/fileadmin/user_upload/CAS_Code_2020__EN_.pdf (last visited Feb 18, 2024).

---

47.  Frequently Asked Questions, (2024), https://www.tas-cas.org/en/general-information/frequently-asked-questions.html (last visited Feb 18, 2024).

48.  A legit supreme court of world sports? The CAS(e) for reform | The International Sports Law Journal, https://link.springer.com/article/10.1007/s40318-021-00184-0 (last visited Feb 18, 2024).

**DIGITAL TRANSFORMATION AND INNOVATION IN SPORTS BUSINESS: OPPORTUNITIES AND CHALLENGES**

Technology is reshaping the sports industry. It posses both challenges and opportunities for sports business administration. However, digital transformation refers to leveraging on digital technologies. It is either used to form new or modify existing business processes. While innovation implies the introduction of novel or enhanced products, services or practices that are valuable for stakeholders, customers or society.[49] The drivers behind the digitization and innovations in sport in business include:

a) The increasing demand and expectation of fans and consumers for personalized, interactive, and immersive experiences across multiple platforms and devices.

b) The emergence and adoption of new technologies such as artificial intelligence (AI), big data analytics, cloud computing, blockchain, internet of things (IoT), 5G, virtual reality (VR), augmented reality (AR), and mixed reality (MR).[50]

c) The growing competition and collaboration among sports organizations, media companies, technology providers, sponsors, and other stakeholders in the sports ecosystem.

d) The changing regulatory and legal environment, especially in relation to data protection, privacy, intellectual property, and antitrust issues.

Some of the major problems associated:

i. Requirement to invest into emerging technology while managing associated risks.

ii. The demand for balancing the needs as well as aspirations of diverse stakeholder segments.

iii. The necessity to safeguard as well as drive revenue from data and IP[51] resulting from sporting activities and events.

iv. Compliance with emerging but intricate legal or moral frameworks that oversee use of digital technology and innovations in sport.[52]

---

49. Digital Transformation and Future Changes in the Sports Industry, DELOITTE UNITED STATES, https://www2.deloitte.com/us/en/pages/technology-media-and-telecommunications/articles/digital-transformation-and-future-changes-in-sports-industry.html (last visited Feb 18, 2024).

50. Explore the Top 8 Sports Industry Trends in 2024 | StartUs Insights, https://www.startus-insights.com/innovators-guide/sports-tech-trends/ (last visited Feb 18, 2024).

51. Intellectual Property.

52. Digital Transformation Is Changing the Face Of Sports?, (Nov. 19, 2021), https://www.brisklogic.co/how-digital-transformation-is-changing-the-face-of-sports/ (last visited Feb 18, 2024).

Some of the main advantages that come with digital transformation and innovation in sports business include:

1.    Opportunity to improve fan experience (through personalized, interactive & immersive experiences) across multiple platforms & devices that will consequently increase customer engagement levels & loyalty.

2.    The utilization of data analytics, AI as well as other technologies to manage training programs, recovery processes and decision making should improve athletes', teams' and trainers' performances and wellbeing.

3.    In addition, the data associated with sports events can be leveraged through monetisation in order to generate streams of revenue that are new or diversified.

4.    In short, digital transformation and innovation are revolutionizing the sports industry in multiple ways; posing challenges and creating opportunities for sports business administration. The changing environment has to be acknowledged by sports business administrators who must also seize these opportunities to create value for their organizations and stakeholders.

Therefore, digital transformation is currently taking place in sport sector leading to new opportunities and threats for managers of sport businesses. To survive in such a context of rapid changes they need not only adapt themselves but also increase organizational values.

## SPORTS RECRUITMENT AND TALENT IDENTIFICATION UNDER DATA-DRIVEN EVALUATION AND AI RECOMMENDATION

Sports organizations depend on sports recruitment and talent identification as the main determinants of human resources quality and future potential.[53] Yet these processes are usually complex, expensive, and subjective because they rely on human judgment, intuition, and bias. They also suffer from information asymmetry whereby not all players have equal access and exposure to scouts and coaches. Consequently, many talented players may be underrated or ignored while many average ones could be overrated or overpaid.

On the other hand, artificial intelligence (AI) can revolutionize sports recruitment and talent identification through data-driven evaluation and recommendation.[54] This is because AI can employ a range of records like performance statistics, visual footages, biometric data sets, psychological profiles as well as scouting reports so as to evaluate different athletes using objective criteria that can be quantified. Besides it utilizes machine learning algorithms as well as computer vision among other techniques in mining the data for insights towards generating predictive models hence ranking players accordingly. AI may then give to the recruiters and coaches, as well as players and agents personalized recommendations which can be acted upon to assist in arriving at a decision and optimize results.[55] Some of the advantages of adopting AI for talents identifying system in sports include:

a. Player evaluation can be made easier, more precise and expedited by AI due to minimization of human errors and biases as well as maximization of data analysis speed, scale, and magnitude.

b. Through AI scouts or coaches are able to reach players from various regions, levels or backgrounds and thus enlarge the pool for talent including all areas which could have been affected by geographic, cultural or social barriers that may hinder the development of the game.

c. The selection process of players can be made more transparent by using artificial intelligence (AI) in order to make it easy for such processes to be assessed. For instance providing objective criteria and metrics which are standardized will promote fair selection techniques while facilitating players and agents in verifying their skills as well-laid potentialities.

d. Tailored feedbacks which are adaptive as well as guidance will help improve some inefficiencies hence enhance performance as well develop actors whereas weaknesses can be exposed through identification of strengths leading betterment.[56]

---

53.    AI in Sports: Use Cases, Examples, and Challenges, https://www.itransition.com/ai/sports (last visited Feb 18, 2024).

54.    The Future of Talent Scouting and Player Analysis: The Impact of AI - Yellowbrick, https://www.yellowbrick.co/blog/sports/the-future-of-talent-scouting-and-player-analysis-the-impact-of-ai (last visited Feb 18, 2024).

AI for sport recruitment and talent identification poses a set of challenges and limitations. Some of them are:

1. Ethical and legal issues can also be raised by AI in terms of data privacy, security, ownership and consent as well as algorithmic accountability, explainability and bias.[57]

2. Technical problems such like availability, integration, reliability, complexity, validity, quality, algorithms, data etc., could also pose a threat to artificial intelligence (AI).

3. Also the social and human aspects including human agency, autonomy and dignity, trust as well as communication among humans can be affected by AI.[58]

---

55. Sports and AI. Eight Uses of AI in Sports | Plat.AI, https://plat.ai/blog/uses-of-artificial-intelligence-in-sports/ (last visited Feb 18, 2024).

56. AI in Sports, *supra* note 46.

57. Nader Chmait & Hans Westerbeek, *Artificial Intelligence and Machine Learning in Sport Research: An Introduction for Non-Data Scientists*, 3 FRONTIERS IN SPORTS AND ACTIVE LIVING (2021), https://www.frontiersin.org/articles/10.3389/fspor.2021.682287 (last visited Feb 18, 2024).

**CONCLUSION**

To sum up AI can improve sports recruitment and talent identification through providing data-driven evaluation and recommendation. However, there are also some challenges that come with the use of AI which should be addressed so that they do not become limiting factors to the technology. Therefore, rather than substituting human judgment or expertise AI instead is meant to be integrated into them by ensuring provision of valuable information or insights that would facilitate their enhancement or complementing them.

.

58.   Kevin Till & Joseph Baker, *Challenges and [Possible] Solutions to Optimizing Talent Identification and Development in Sport*, 11 FRONTIERS IN PSYCHOLOGY (2020), https://www.frontiersin.org/journals/psychology/articles/10.3389/fpsyg.2020.00664 (last visited Feb 18, 2024).